\documentclass[aps,prl,twocolumn,floats,epsfig,showpacs]{revtex4-1}
\usepackage{bbm}
\usepackage{amssymb}
\usepackage{ifpdf}
\usepackage{graphicx,epsfig}
\newcommand{\Z}{{\mathbb{Z}}}

\begin{document}

\hspace*{0.75\textwidth}\texttt{\footnotesize CERN-PH-TH/2012-091}

\title{Walking near a Conformal Fixed Point: the 2-d $O(3)$ Model at 
$\theta \approx \pi$ as a Test Case}

\author{P.\ de Forcrand$^{a,b}$ , M.\ Pepe$^c$, and U.-J.\ Wiese$^d$}

\affiliation{
$^a$ Institute for Theoretical Physics, ETH Z\"urich, CH-8093 Z\"urich,
Switzerland \\
$^b$ CERN, Physics Department, TH Unit, CH-1211 Gen\`eve 23, Switzerland \\
$^c$ INFN, Istituto Nazionale di Fisica Nucleare, Sezione di Milano-Bicocca
Edificio U2, Piazza della Scienza 3, 20126 Milano, Italy \\
$^d$ Albert Einstein Center for Fundamental Physics, Institute for Theoretical 
Physics, Bern University, Sidlerstrasse 5, CH-3012 Bern, Switzerland}

\begin{abstract}
Slowly walking technicolor models provide a mechanism for electroweak symmetry 
breaking whose nonperturbative lattice investigation is rather challenging. Here
we demonstrate walking near a conformal fixed point considering the 2-d lattice 
$O(3)$ model at vacuum angle $\theta \approx \pi$. 
The essential features of walking technicolor models are shared by this
toy model and can be accurately investigated by numerical simulations. We
show results for the running coupling and the beta-function and we perform
a finite size scaling analysis of the massgap close to the conformal
point.
\end{abstract} 

\maketitle

The scalar Higgs field that drives spontaneous electroweak symmetry breaking in 
the Standard Model is considered unnatural as a fundamental degree of freedom
because it suffers from the gauge hierarchy problem. Technicolor provides a 
promising mechanism that stabilizes the electroweak scale against the Planck or
GUT scale by introducing a new asymptotically free strongly coupled gauge theory
\cite{Hil02}. The chiral condensate of techniquarks then induces electroweak 
symmetry breaking and replaces the fundamental Higgs field in a natural way, 
i.e.\ without fine-tuning. The original technicolor models 
\cite{Wei76,Sus79,Far79} suffer from flavor-changing neutral currents. In 
addition, electroweak precision tests are strongly affected by the running of 
the technicolor gauge coupling \cite{Pes90}. In order to avoid these problems, 
it has been proposed to consider slowly walking technicolor theories 
\cite{Hol81,App87,Die05}. In these models the running of the gauge coupling 
slows down due to the proximity to a conformal fixed point \cite{Ban82}. Besides
non-Abelian technicolor gauge fields, these models usually contain many flavors 
of techniquarks in the fundamental representation or technifermions in 
higher-dimensional representations of the technicolor gauge group. In order to 
naturally stabilize the electroweak scale against the Planck scale, these 
models are still asymptotically free. They are just outside the conformal 
window of theories that have an infrared conformal and an ultraviolet 
asymptotically free fixed point. The lower edge of the conformal window 
corresponds to the merging of two fixed points and thus to a double zero of the 
$\beta$-function \cite{Gie06,Kap09}. Determining whether a theory is inside or 
near the conformal 
window requires nonperturbative investigations, which can be performed from 
first principles using lattice gauge theory. Lattice theories with different 
gauge groups and with different fermionic matter content have been studied 
quite intensively \cite{App08,Deu08,Hie09,Fod09,Del10,DeG10,Has12}. Due to 
finite-size and finite lattice spacing effects as well as due to difficulties 
with simulating nearly massless dynamical fermions, investigating whether a 
theory is inside the conformal window is a difficult task. Determining whether 
a theory is slowly walking is even more challenging. At present, there is no
consensus on which theories are slowly walking.

In this paper, for the first time we unambiguously demonstrate walking near a
conformal fixed point in an asymptotically free field theory --- the 2-d $O(3)$
model, which shares many features with 4-d non-Abelian gauge theories. In
particular, besides being asymptotically free, it has a dynamically generated
massgap, instantons, and thus a non-trivial vacuum angle $\theta$. At 
$\theta = 0$, the Euclidean action of the model is given by
\begin{equation}
S[\vec e] = \frac{1}{2 g^2} 
\int d^2x \ \partial_\mu \vec e \cdot \partial_\mu \vec e.
\end{equation}
Here $\vec e(x)$ is a 3-component unit-vector field and $g$ is the coupling
constant. The topological charge,
\begin{equation}
Q[\vec e] = \frac{1}{8 \pi }\int d^2x \ \varepsilon_{\mu\nu} \vec e \cdot
(\partial_\mu \vec e \times \partial_\nu \vec e) \in \Pi_2[S^2] = \Z,
\end{equation}
contributes an additional term $i \theta Q[\vec e]$ to the action.
The 2-d $O(3)$ model is integrable at $\theta = 0$ \cite{Zam79,Pol83,Wie85} as 
well as at $\theta = \pi$ \cite{Zam86}, but not at intermediate values of 
$\theta$. 
The massgap at $\theta = 0$ has been determined analytically \cite{Has90} and is
given by $M = \frac{8}{e} \Lambda_{\overline{MS}}$, where $e$ is the base of the 
natural logarithm and $\Lambda_{\overline{MS}}$ is the scale that is dynamically 
generated by dimensional transmutation in the modified minimal subtraction 
scheme. As one varies $\theta$, the massgap is reduced until it finally vanishes
at $\theta = \pi$. The conjectured exact S-matrix at $\theta = \pi$ has 
recently been confirmed by lattice simulations with per mille level accuracy 
\cite{Boe11}. This study has also demonstrated beyond any reasonable doubt that
$\theta$ is a relevant parameter that does not get renormalized 
nonperturbatively. Consequently, each value of $\theta$ characterizes a
different physical theory, which was further supported by \cite{Nog12}.

As one infers from the exact S-matrix \cite{Zam86}, at energies far below 
$\Lambda_{\overline{MS}}$ the 2-d $O(3)$ model at $\theta = \pi$ reduces to the 
$k = 1$ WZNW model \cite{Wes71,Nov81,Wit84}, a conformal field theory with the 
Euclidean action
\begin{equation}
S[U] = \frac{1}{2 {g'}^2} \int d^2x \
\mbox{Tr}[\partial_\mu U^\dagger \partial_\mu U] - 2 \pi i k S_{WZNW}[U].
\end{equation}

\begin{figure}[t]
\vskip-0.05cm
\includegraphics[width=0.48\textwidth]{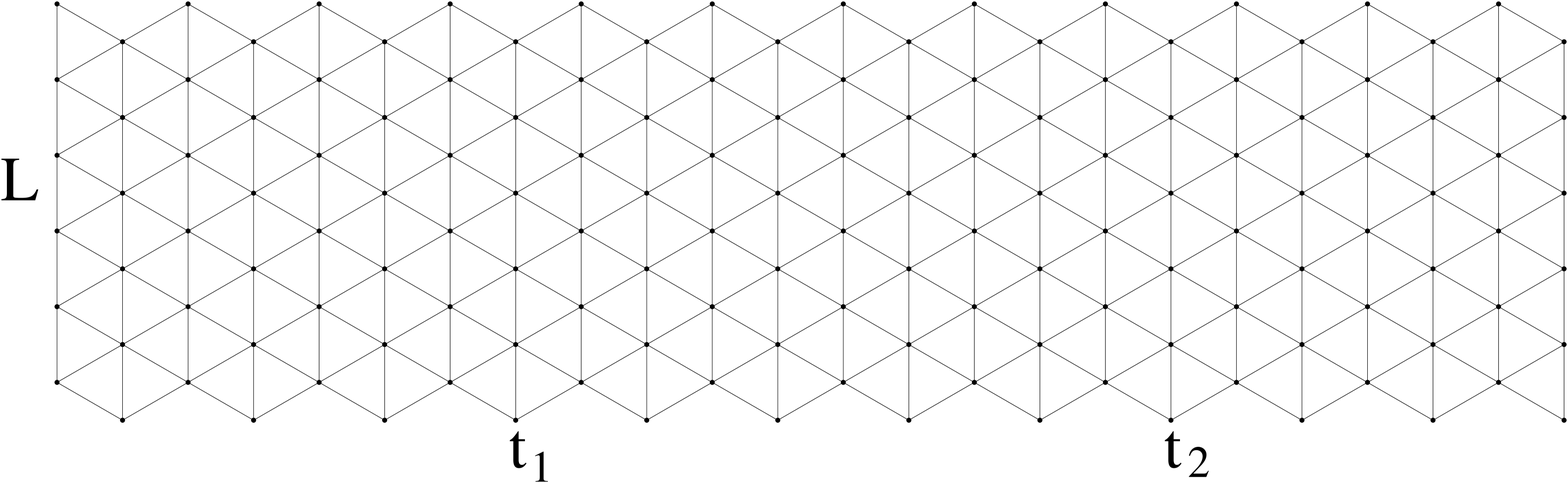}
\vskip-0.5cm
\caption{\it Triangular lattice of size $L \times \beta$. The triangles 
$\langle xyz \rangle$ carry the topological term $i \theta q_{\langle xyz \rangle}$.
The correlation function $C(t_1,t_2;\theta)$ is determined between the
time-slices at $t_1$ and $t_2$.}
\end{figure}

Here $U(x) \in SU(2) = S^3$ and $g'$ is a coupling constant. 
The Wess-Zumino-Novikov-Witten term is given by
\begin{eqnarray}
&&S_{WZNW}[U] = \nonumber \\
&&\frac{1}{24 \pi^2} \int_{H^3} d^2x \ dx_3 \
\varepsilon_{\mu\nu\rho} \mbox{Tr}
[U^\dagger \partial_\mu U U^\dagger \partial_\nu U U^\dagger \partial_\rho U].
\end{eqnarray}
Here $H^3$ is a 3-d hemisphere whose boundary $\partial H^3 = S^2$ is the 
compactified 2-d space-time. Since $\Pi_2[S^3] = \{0\}$, there are no 
topological obstructions against extending the 2-d field $U(x)$ on $x \in S^2$ 
to a 3-d field $U(x,x_3)$ on $(x,x_3) \in H^3$. The WZNW-terms corresponding to 
two different extensions $U^{(1)}$ and $U^{(2)}$ differ by an integer,
\begin{eqnarray}
&&S_{WZNW}[U^{(1)}] - S_{WZNW}[U^{(2)}] = \frac{1}{24 \pi^2} 
\int_{S^3} d^2x \ dx_3 \nonumber \\
&&\times \varepsilon_{\mu\nu\rho} \mbox{Tr}
[U^\dagger \partial_\mu U U^\dagger \partial_\nu U U^\dagger \partial_\rho U]
\in \Pi_3[S^3] = \Z.
\end{eqnarray}
Here two hemispheres have been combined to form a sphere $S^3$ with
$U$ corresponding to $U^{(1)}$ on one and to $U^{(2)}$ on the other hemisphere. 
Since $S_{WZNW}[U]$ is thus well-defined only up to an integer, in order to 
obtain an unambiguous contribution $\exp(2 \pi i k S_{WZNW}[U])$ to the 
functional integral, the level $k$ must be quantized in integer units. 
Interestingly, the WZNW model has a global $SU(2)_L \times SU(2)_R$ symmetry, 
$U(x)' = L U(x) R^\dagger$, which extends the $O(3)$ symmetry to $O(4)$. It 
should be noted that only the low-energy WZNW sector, but not the entire $O(3)$ 
model at $\theta = \pi$ has the enlarged $O(4)$ symmetry. The scale 
$\Lambda_{\overline{MS}}$, which results from anomalous scale breaking and
dimensional transmutation, still induces explicit symmetry breaking down to 
$O(3)$. Due to the enlarged $O(4)$ symmetry of the low-energy sector, an 
$O(3)$ singlet becomes degenerate with the $O(3)$ triplet as 
$\theta \to \pi$ \cite{Con04}.

Since the WZNW model is a conformal field theory, the 2-d $O(3)$ model at 
$\theta \approx \pi$ is a natural candidate for a slowly walking asymptotically 
free theory near a conformal fixed point \cite{Nog12}. Following 
\cite{Lue82,Lue91}, we define a running coupling constant 
$\alpha(\theta,L) = g^2(\theta,L) \equiv m(\theta,L) L$ through the massgap $m(\theta,L)$ in a periodic 
volume of spatial size $L$. At small $L$ the coupling $\alpha(\theta,L)$ can be 
computed in perturbation theory. In this limit, it is independent of the vacuum 
angle $\theta$, which does not affect perturbation theory, and it agrees with 
the standard asymptotically free coupling constant of the 2-d $O(3)$ model. For
large $L$, on the other hand, the coupling $\alpha(\theta,L)$ is $\theta$-dependent
and can only be computed nonperturbatively. The corresponding $\beta$-function 
is given by $\beta(\theta,\alpha) = - L \partial_L \alpha(\theta,L)$. Thanks to an 
ingenious use of the thermodynamic Bethe ansatz the finite-volume massgap is 
known analytically, both at $\theta = 0$ \cite{Bal04} and at $\theta = \pi$ 
\cite{Bal11}. Although the 2-d $O(3)$ model is not integrable for other values 
of $\theta$, by expanding around $\theta = \pi$ some interesting analytic 
results have been obtained even in that regime \cite{Con04}. Due to the 
non-vanishing massgap $M$ at $\theta = 0$, for large $L$ the coupling 
$\alpha(0,L) \rightarrow M L$ increases linearly with the volume. At $\theta = \pi$, 
on the other hand, the massgap vanishes and the coupling approaches a fixed 
point $\alpha(\pi,L) \rightarrow \alpha^\star=\pi$ as one increases $L \rightarrow \infty$. 
In
addition to the Gaussian fixed point at $\alpha = 0$, the $\beta$-function has
another zero at $\alpha^\star = \pi$, i.e.\ $\beta(\theta = \pi,\alpha = \pi) = 0$ \cite{Aff89}.
Since the physics is symmetric
around $\theta = \pi$, at the fixed point the $\beta$-function just touches,
but does not cross zero. The double zero of the
$\beta$-function thus corresponds to the lower edge of the conformal window.
The parabolic form of the $\beta$-function near this fixed point causes
large logarithmic finite-size corrections:
\begin{equation}
\beta(\alpha) \approx -C (\alpha - \alpha^\star)^2 ~\Rightarrow~
\alpha(L) \approx \alpha^\star - \frac{1}{C \log(L/L_0)}
\label{eq:logcorr}
\end{equation}
These logarithmic corrections lead to a very slow approach to the conformal 
fixed point.
In our model, they are associated with a marginally irrelevant operator that 
breaks the $O(4)$ symmetry down to $O(3)$ \cite{Car86,Aff87,Aff89,Rea90}.

Besides the running or walking coupling, the $\theta$-dependence of the 
infinite-volume massgap is also of interest. Near the fixed point at 
$\theta = \pi$ it is given by \cite{Aff89}
\begin{equation}
\label{scaling}
m(\theta,L \rightarrow \infty) \sim 
|\theta - \pi|^{2/3} |\log(|\theta - \pi|)|^{-1/2}.
\end{equation}

The determination of $m(\theta,L)$ and $\alpha(\theta,L)$ requires 
nonperturbative calculations. These can be performed from first principles 
using lattice field theory. Simulating the 2-d lattice $O(3)$ model at large 
$\theta$ is very challenging due to a severe sign problem. Fortunately, the 
meron-cluster algorithm \cite{Bie95}, which is based on the Wolff cluster 
algorithm \cite{Wol89}, substantially reduces the sign problem and thus makes a 
numerical study feasible. For technical reasons related to the meron-cluster 
algorithm, we consider the 2-d $O(3)$ model on a triangular lattice of spatial 
extent $L$ and Euclidean time extent $\beta > L$ as illustrated in Figure 1. 
The action is defined on nearest-neighbor bonds 
$\langle xy \rangle$, and is given by 
\begin{equation}
S[\vec e] = \sum_{\langle xy \rangle} s(\vec e_x,\vec e_y), \quad
s(\vec e_x,\vec e_y) = \frac{1}{g^2} (1 - \vec e_x \cdot \vec e_y),
\end{equation} 
for $\vec e_x \cdot \vec e_y > - \frac{1}{2}$ and 
$s(\vec e_x,\vec e_y) = \infty$ otherwise. This action eliminates field 
configurations for which the angle between neighboring spins exceeds 120 
degrees, which is essential for the success of the meron-cluster algorithm. The 
geometric topological charge density 
$q_{\langle xyz \rangle} \in [- \frac{1}{2},\frac{1}{2}]$  \cite{Ber81} associated 
with a triangle $\langle xyz \rangle$ is given by
\begin{eqnarray}
R \exp(2 \pi i q_{\langle xyz \rangle})&=&1 + \vec e_x \cdot \vec e_y +
\vec e_y \cdot \vec e_z + \vec e_z \cdot \vec e_x \nonumber \\
&+&i \vec e_x \cdot (\vec e_y \times \vec e_z), \quad R \geq 0.
\end{eqnarray}
Here $4 \pi q_{\langle xyz \rangle}$ is the oriented area of the spherical triangle 
on $S^2$ defined by the three unit-vectors $\vec e_x$, $\vec e_y$, and 
$\vec e_z$. By construction, the geometric lattice topological charge 
$Q[\vec e] = \sum_{\langle xyz \rangle} q_{\langle xyz \rangle}$ is an integer. In order
to determine the massgap, we consider the operator 
$\vec E(t) = \sum_{x_1} \vec e_x$, where the sum extends over all points
$x = (x_1,t)$ in a time-slice, and we define the 2-point function
\begin{eqnarray}
&&C(t_1,t_2;\theta) = \frac{1}{Z(\theta)} \prod_x \int_{S^2} d\vec e_x \ 
\vec E(t_1) \cdot \vec E(t_2) \nonumber \\
&&\times \exp(- S[\vec e] + i \theta Q[\vec e]) \sim
\exp(- m(\theta,L)(t_2-t_1)), \nonumber \\ 
&&Z(\theta) = \prod_x \int_{S^2} d\vec e_x \
\exp(- S[\vec e] + i \theta Q[\vec e]),
\end{eqnarray}
which decays exponentially with the $\theta$- and $L$-dependent 
massgap $m(\theta,L)$ at large Euclidean time separations. 

Like in the Wolff cluster algorithm, in the meron-cluster algorithm one first
chooses a reflection plane perpendicular to a randomly selected unit-vector
$\vec r \in S^2$. In a given update step, the spins $\vec e_x$ are either left
unchanged, or they are reflected to 
$\vec e_x~\!\!\!'  = \vec e_x - 2 (\vec e_x \cdot \vec r) \vec r$. Spins belonging to
a common cluster are reflected collectively \cite{Wol89}. In the meron-cluster
algorithm, each cluster ${\cal C}$ contributes an integer or half-integer to 
the total topological charge $Q$. Thanks to the 120 degrees angle constraint, 
the cluster topological charge $Q_{\cal C}$ is well-defined and does not depend 
on whether other clusters are flipped or not. Clusters with a half-integer 
topological charge are called
meron-clusters, because they represent half-instantons. Flipping a meron-cluster
changes the topological charge by an odd integer. At $\theta = \pi$ where
$\exp(i \theta Q) = (-1)^Q$, this implies an exact cancellation between two
contributions to the partition function. Taking this cancellation into account
analytically rather than statistically via Monte Carlo leads to an improved 
estimator that exponentially reduces the sign problem. In a multi-cluster 
algorithm, for general values of $\theta$ an improved estimator for the 
partition function is given by
\begin{equation}
\frac{Z(\theta)}{Z(0)} = \langle \prod_{\cal C} \cos(\theta Q_{\cal C}) \rangle.
\end{equation}
Similarly, for the 2-point function one obtains
\begin{eqnarray}
&&\langle \vec e_x \cdot \vec e_y \exp(i \theta Q) \rangle = \langle
(\vec e_x \cdot \vec r)(\vec e_y \cdot \vec r) 
\prod_{\cal C} \cos(\theta Q_{\cal C}) \nonumber \\
&&\times \left[ \delta_{{\cal C}_x,{\cal C}_y} + (1 - \delta_{{\cal C}_x,{\cal C}_y})
\tan(\theta Q_{{\cal C}_x}) \tan(\theta Q_{{\cal C}_y}) \right]\rangle,
\end{eqnarray}
where the cluster ${\cal C}_x$ contains $x$ and ${\cal C}_y$ contains $y$.

\begin{figure}[t]
\begin{center}
\includegraphics[width=0.50\textwidth,angle=0]{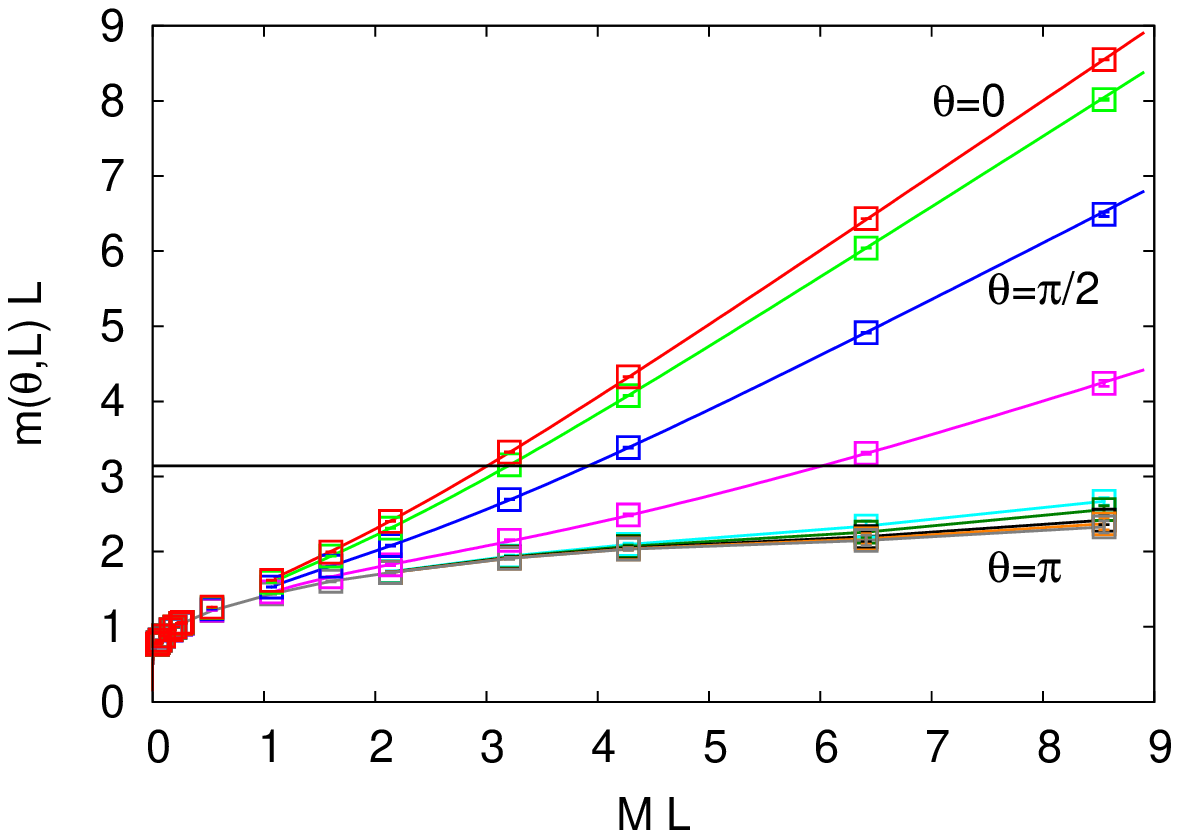} \\
\includegraphics[width=0.50\textwidth,angle=0]{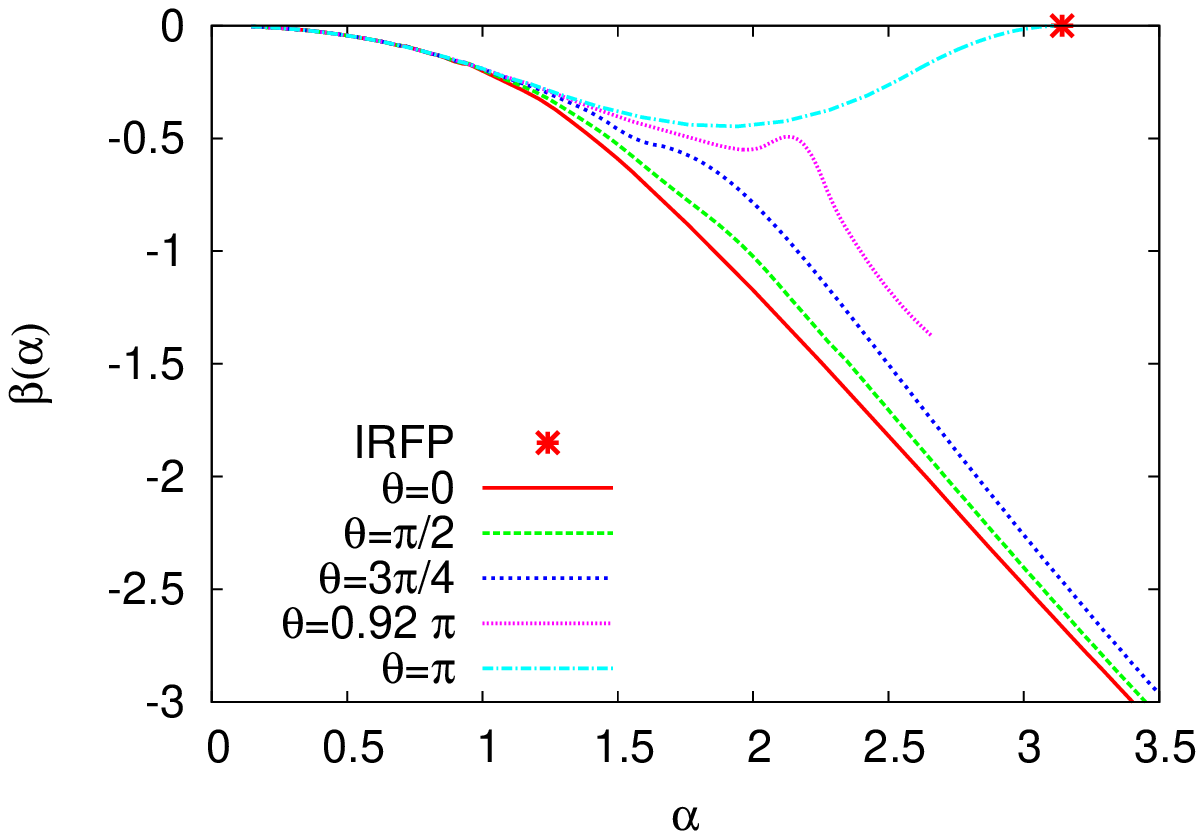} \\
\caption{\it (top) The running coupling constant $\alpha(\theta,L)=m(\theta,L) L$ as a function of the 
scale set by the spatial size $L$ for $\theta/\pi = 0, 0.25, 0.50, 0.75, 0.92,
0.94, 0.96, 0.98$ and $1$.  At $\theta \neq \pi$ 
the coupling increases linearly as $L \rightarrow \infty$, while at 
$\theta = \pi$ it approaches the fixed point value $\pi$ (indicated by the 
horizontal line). At $\theta \approx \pi$ the coupling is slowly walking 
near the conformal fixed point. The curves for $0\leq \theta \leq 3\pi/4$ show the effect of leading 
finite-size corrections $\exp(-mL)/\sqrt{mL}$ to the infinite-size massgap $m$.
(bottom) The corresponding $\beta$-function obtained by differentiation of our
numerical results, supplemented by exact values at $\theta=\pi$ from \cite{Bal11}.}
\end{center}
\end{figure}

We have used the meron-cluster algorithm to determine the massgap $m(\theta,L)$
for a large variety of $\theta$-values and volumes $L$ ranging from 6 to 100. 
By using at least four values of the bare coupling $g$, all results have 
been reliably extrapolated to the continuum limit. The corresponding results for
the running coupling $\alpha(\theta,L)$ are illustrated at the top of Figure 2. 
Within error bars, they agree with values obtained from the exact massgap
both at $\theta = 0$ and at $\theta = \pi$ \cite{Bal11}.
While $\alpha(\theta,L)$ increases linearly for large $L$ when 
$\theta \neq \pi$, it flattens off for $\theta = \pi$. As anticipated, due to
large logarithmic corrections the approach to the fixed point is very slow. 
For $\theta \approx \pi$, the coupling walks 
almost as slowly as at $\theta = \pi$ up to some distance scale, at which it 
starts running off into the linearly rising regime. While in a lattice context
it is most natural to use a step scaling function \cite{Lue91}, here we prefer
to discuss the $\beta$-function that is familiar from the continuum, although 
its computation requires a spline-interpolation of the lattice data. The 
walking versus running of the coupling manifests itself in the $\beta$-function 
shown at the bottom of Figure 2. For $\theta \approx \pi$ it walks 
towards the fixed point $\beta(\theta = \pi,\alpha = \pi) = 0$ before running off (linearly) to 
negative values.

Figure 3 shows the finite-size scaling behavior of 
$m(\theta,L) L + [\pi - m(\pi,L) L]$ as
a function of $M L t^{2/3} / \sqrt{|\log(t/t_0)|}$, 
where $t$ is the reduced coupling $t=1-\theta/\pi$, and $t_0=70$. Large
logarithmic corrections are removed in this difference between the 
finite-volume massgaps, so that the $L\to\infty$ value $\pi$ is enforced
at the critical coupling $t=0$. 
The fact that all data fall on a universal curve confirms the behavior
of the massgap eq.(\ref{scaling}), and in particular the critical
exponent $2/3$ of the WZNW model.

\begin{figure}[b]
\label{massgap}
\begin{center}
\includegraphics[width=0.48\textwidth,angle=0]{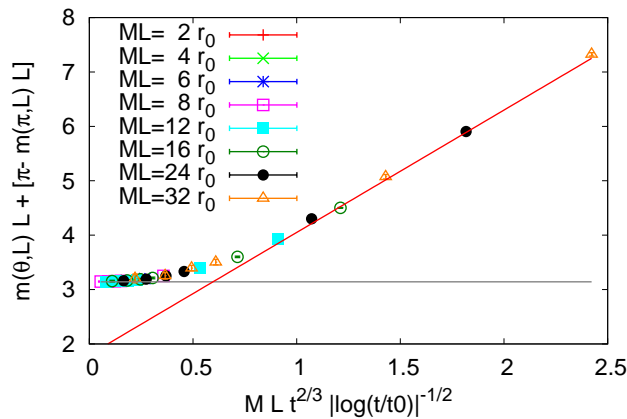}
\caption{\it Finite-size scaling of $m(\theta,L) L$ near 
$\theta = \pi$ as a function of 
$M L t^{2/3} / \sqrt{|\log(t/t_0)|}$, with $t=1-\theta/\pi, t_0=70$. 
System sizes $L$ have been chosen as multiples of $r_0/M$, with
$r_0=0.26715356$ \cite{Bal10}.
The data have been shifted by $[\pi - m(\pi,L)L]$, to eliminate the large
logarithmic corrections and enforce $m(\theta=\pi,L)L = \pi$. They fall on a 
universal curve, confirming the critical exponent $2/3$ of the WZNW model in 
eq.(\ref{scaling}). This curve shows both the slow walking near $\pi$, and
the linearly rising regime as $L\to\infty$.
}
\end{center}
\end{figure}

It is interesting to compare the behavior of the 2-d $O(3)$ model near the 
conformal fixed point at $\theta = \pi$ with the anticipated behavior of 
walking 4-d non-Abelian technicolor gauge theories near the conformal window
\cite{Nog12}. Both theories are asymptotically free and conformality is 
thus limited to scales far below $\Lambda_{\overline{MS}}$.
$(i)$ First of all, in the 2-d $O(3)$ model the parameter that 
determines the distance to the conformal fixed point is the continuously 
varying vacuum angle $\theta$, which does not get renormalized 
\cite{Boe11,Nog12}, and which does not affect the $\beta$-function in the 
perturbative regime. In walking technicolor theories, on the other hand, the 
corresponding parameter is the discrete number of techniquark flavors or the 
size of the technifermion representation, which do affect the perturbative 
$\beta$-function. Unlike in the 2-d $O(3)$ model, due to renormalization, one 
can then not directly compare physical quantities between theories in and 
outside the conformal window. $(ii)$ While large, logarithmic finite-size effects
are a characteristic of walking theories near the edge of the conformal window,
as shown by eq.(\ref{eq:logcorr}), their origin may differ. As we have seen,
in the 2-d $O(3)$ model a marginally irrelevant operator breaks the enhanced
$O(4)$ symmetry in the low-energy sector. In a 4-d non-Abelian technicolor 
gauge theory, it is not clear whether logarithmic corrections could come from
a similar symmetry enhancement, or from conformal symmetry itself.
$(iii)$ In the 2-d $O(3)$ model the $O(4)$ symmetry enhancement causes an
$O(3)$ singlet to become light as $\theta\to\pi$, in addition to the $O(3)$ triplet, all having
masses much below $\Lambda_{\overline{MS}}$, before both objects 
become massless ``unparticles'' \cite{Geo07} at $\theta = \pi$. It has been 
argued that particles with a mass 
much below $\Lambda_{\overline{MS}}$ should also arise in walking technicolor 
theories near the edge of the conformal window \cite{Yam86,Has11}. These
so-called technidilatons have been identified with pseudo-Nambu-Goldstone 
bosons of a spontaneously broken conformal invariance, which is still weakly 
explicitly broken by the scale anomaly at $\Lambda_{\overline{MS}}$. In the 2-d 
\linebreak
$O(3)$ model, due to the Mermin-Wagner theorem, conformal invariance cannot 
break spontaneously and thus the light $O(3)$ triplet and singlet are not 
expected to be pseudo-Nambu-Goldstone bosons. In particular, they are exactly
massless at $\theta = \pi$, despite the fact that conformal symmetry exists only
in the low-energy sector and is explicitly broken at the scale 
$\Lambda_{\overline{MS}}$. 
$(iv)$ In addition to the large finite-size effects, cut-off effects
due to a finite lattice spacing $a$ may also be important.
In the 2-d $O(3)$ model, large logarithmic corrections to the expected 
${\cal O}(a^2)$ effects mimic ${\cal O}(a)$ behavior \cite{Bal10}. 
Similarly, in lattice investigations of technicolor gauge theories, 
one must control these cut-off effects. When one uses Wilson fermions, 
which are theoretically cleaner than staggered fermions, without Symanzik 
improvement lattice artifacts are of order $a$. 
$(v)$ Finally, in 
a non-Abelian gauge theory the Schr\"odinger functional \cite{Lue92} provides a 
definition of a running or walking coupling constant, which naturally replaces 
the coupling based on the finite-volume massgap that we use in the 2-d $O(3)$ 
model. 

In the end, 
our study of the 2-d $O(3)$ model near $\theta = \pi$ demonstrates 
that slow walking can indeed be studied accurately using Monte Carlo
simulations, provided that lattice artifacts and finite-volume effects, which 
may both be large, are well understood and under good numerical control. 
Besides technicolor gauge theories, it would be interesting to also investigate 
other models, e.g.\ 4-d Yang-Mills theories at non-zero $\theta$, in order to 
further investigate the neighborhood of conformal fixed points.

We are indebted to J.\ Balog, M.\ Blau, C.\ Destri, and E.\ Vicari for 
illuminating discussions. P.\ de F.\ thanks YITP, Kyoto, for its hospitality,
and M.\ P.\ acknowledges P.\ Ramieri for technical support. This work has 
been supported by the Regione Lombardia and CILEA Consortium through a LISA 2011
grant, as well as by the Schweizerischer Nationalfonds (SNF).

\end{document}